\documentclass[aps,pra,superscriptaddress,preprint,showpacs]{revtex4}

\usepackage{graphicx}
\usepackage{graphics}
\usepackage{amssymb}
\usepackage{amsmath}
\usepackage{epsfig}
\usepackage{latexsym}
\usepackage{color}
\usepackage{rotating}
\usepackage{subfigure}

\begin{document}

\title{A quantum optomechanical Mach clock.}

\author{G. J. Milburn}
\email{milburn@physics.uq.edu.au}
\affiliation{Centre for Engineered Quantum Systems, School of Mathematics and Physics, The University of Queensland, St Lucia, QLD 4072, Australia}
\author{T. J. Milburn}

\begin{abstract}
We introduce a quantum thermodynamic system that can operate as a thermal clock in analogy with the
thermal clock first introduced by Ernst Mach in which the temperature difference between cooling bodies in contact can be used as a kind of relational time.  Our model is based on an optomechanical system in which photons are transferred irreversibly between two cavities due to  the modulation of the cavity cooling rate by a mechanical system coupled to a heat bath. We contrast the ensemble average view with a single system view using the theory of continuous measurement and we show that, by making a weak continuous measurement of the energy difference between the systems, a natural Mach thermal clock arises  in an appropriate semiclassical limit. We investigate how quantum fluctuations modify this result.
\end{abstract}

\maketitle
\section{Introduction}

Mach introduced a temperature clock in an attempt to ground an understanding of time in our sense perception of irreversible processes~\cite{Mach1}.  Mach's clock is a thermodynamic system in which three identical materials, initially held at different temperatures, are placed in thermal contact but thermally isolated from the rest of the universe.  Using Newton's law of cooling, Mach showed that we can tell the time by using a thermometer.  Mach's temperature clock is non-periodic in contrast with the prototypical periodic mechanical clock---an astronomical clock---in which time is told by measurement of an angular coordinate.  

Periodic clocks are more familiar, yet non-periodic clocks are equally good for coordinating kinematics~\cite{BraMil}, and in fact a particular irreversible clock is in common usage: radiocarbon dating~\cite{fn:allClocksIrr}. Non periodic clocks seem to depend upon irreversibility in an essential way unlike periodic  clocks. In reality all mechanical clocks are irreversible machines operating on a stable limit cycle\cite{Denny} and are thus subject to phase fluctuations. Mach himself was well aware of this. As he says in Knowledge and Error, \textit{Indeed, if
we look at these processes carefully and not just schematically, then like
all kinds of oscillations they are not strictly periodic but contain irreversible
components.}\cite{Mach1}   

In this paper we consider a quantum model for Mach's temperature clock in order to elucidate the role of quantum flucutations. For such a model, temperature is not the focus of attention but rather we must necessarily consider the role of fluctuations in continuously measured quantities of a single realisation of the clock.  There is currently a great deal of interest in how classical thermodynamic concepts should be recast in order to take account of quantum fluctuations~\cite{Jarzynski, Crooks, Hanggi, Vedral, Pekola}. Recently the role of a clocks in quantum thermodynamic engines has been highlighted\cite{Frenzel}.  

Our approach will be based on quantum analogs for Newton's law of cooling:  that the rate of change of the temperature of a body is proportional to the temperature difference between the body and its ambient temperature~\cite{Winterton}.  Strictly speaking this is not a fundamental law, but an empirical observation valid only under rather specific circumstances (e.g., ignoring radiative cooling).  Newton used this observation to define a temperature scale in terms of temporal measurements~\cite{Besson}.  This is precisely the converse to Mach's approach.  That is, Newton uses a clock to define a temperature scale whereas Mach uses a thermometer to define a temporal scale---this duality is in fact quite generic~\cite{Rovelli, Menicucci}.  By giving both a single-system and ensemble picture of Newton's law of cooling, our approach also provides a perspective on Newton's law of cooling in terms of fluctuations.

The paper is structured as follows. In section \ref{thermal-clock} we generalise the non-periodic, irreversible clock based on spontaneous decay of a two-state system to the stochastic transitions between ground and excited states when a two-level system is in thermal equilibrium with a bath at temperature $T$. This enables us to connect temperature and time with quantum fluctuations.
 
In section \ref{qubit-cooling} we give an abstract model for a Newton-like cooling law based on two qubits subject to a stochastic interaction.  This model enables us to introduce the key role played by weak continuous measurement of energy to distinguish ensemble dynamics from stochastic dynamics of a single system. The measurement record now plays the role that measurement of temperature plays in Mach's construct. 

In the main part of the paper, we describe a more physically relevant model based  on an opto-mechanical system in which the exchange of photons between two optical cavities is modulated by coupling to a mechanical degree of freedom. The mechanical system is coupled to a heat bath which maintains it at thermodynamic equilibrium. 

In order to realise an analogue for Mach's thermal clock with only a single instantiation of the optomechanical system, we introduce weak continuous measurements of the photon number difference between the cavities.  This is equivalent to a continuous measurement of the energy of each cavity.  Quantum deviations from semi-classical behaviour can be attributed to collective quantum coherence features. Using an analogy with Dicke superradiance we show how quantum fluctuaitons in our optomehanical example depart from the thermodynmic Mack clock.  


\section{Non periodic clocks.}
\label{thermal-clock}

At first sight the notion of a non periodic, irreversible clock seems paradoxical:  a clock is the very epitome of a deterministic periodic system.  Yet irreversible clocks have long been used, water clocks for  example, and a particular irreversible clock is in common usage.  Radiocarbon dating is based on the stochastic and irreversible decay of a radionuclide---$\text{C}^{14}$. The key point is that metabolism ensures living organisms will continually refresh their concentration of $\text{C}^{14}$ while alive, but once dead the concentration begins to relax to the steady state non-organic concentration of $\text{C}^{14}$ in the environment. As we will see, relaxation from a non equilibrium state is a key feature of non deterministic clocks, especially when the non equlibrium state is a result of state conditioning of a system subject to measurement.  Indeed, in radiocarbon dating one must take into account that different organisms die with differing concentations of $\text{C}^{14}$. Furthermore, the vital concentration may change with time due to concentration of $\text{C}^{14}$ in the atmosphere changing over time. In other words, different organisms and different epochs do not necessarily begin with the same non equilibrium distribution.  

It is known that the probability of a single radionuclide not to decay in a time $t$ is $\exp(- \gamma t)$, where $\gamma$ is called the decay rate.  Given an ensemble of radionuclides, the number that have decayed after some time $t$ is therefore a Poisson distributed stochastic variable $N$ with parameter $\gamma t$.  The mean of $N$ is then simply $\gamma t$.  Thus, given a count of the number of radionuclides that have decayed, $N$, we may estimate the time that has elapsed as
\begin{equation}
	t_{\text{est.}} = \frac{N}{\gamma} .\label{eq:radiocarbonEstimate}
\end{equation}
The error in this estimate is given by the fluctuations in the final count $\delta N$ for a fixed estimate $t_{\text{est.}}$.  Since for a Poisson process the variance is equal to the mean, one finds the relative error is
\begin{equation}
	\frac{\delta t_{\text{est.}}}{t_{\text{est.}}} = \frac{1}{\sqrt{\gamma t_{\text{est.}}}} .\label{eq:radiocabonError}
\end{equation}
Conveniently, we get a better estimate the longer we wait.  Now, in the case of radio-carbon dating, one indeed wishes to compute the elapsed time $t_{\text{est.}}$, and so one requires to know $\gamma$ before hand; calibrate the clock.  However, conversely, one could instead take this system as defining a temporal scale, where to take the time would be to make the count $N$, and the value of $\gamma$ would merely reflect a particular choice of units and thus be purely conventional.  That is, the actual count $N$, while subject to fluctuations, is a local physical quantity that can serve as physical time.  Yet it is worth noting that a radionuclide can only decay, i.e., $N$ can only increase, so in contrast to the prototypical reversible clock, this is a clock with no tock.

A simple generalisation that introduces temperature in a fundamental way results if we consider a two-level system in a thermal bath at temperature $T$ (radiative damping).  While the ensemble average is a stationary state in equilibrium with the bath, continuous measurement of the energy of a single realisation (discussed in more detail below) yields a random telegraph signal.  Let us define the random telegraph signal $Z(t)$ to take the value $Z(t) = -1$ if the two-level system is found in the ground state $|g\rangle$ and $Z(t) = 1$ if it is found in the excited state $|e\rangle$.  The conditional state given this measurement record is then
\begin{equation}
	\rho_{\text{cond.}}(t) = \frac{1 + Z(t)}{2} |e\rangle \langle e| + \frac{1 - Z(t)}{2} |g\rangle \langle g| .
\end{equation}
On the other hand, the ensemble average---the average over all possible random telegraph signals---yields the thermal state
\begin{equation}
	\rho_{\text{th.}}(\beta) = \frac{1 + \tanh(- \beta \epsilon / 2)}{2} |e\rangle \langle e| + \frac{1 - \tanh(- \beta \epsilon / 2)}{2} |g\rangle \langle g| \label{eq:thermalState}
\end{equation}
where $\beta = 1/T$ is the inverse temperature of the bath and $\epsilon$ is the energy separation between $|g\rangle$ and $|e\rangle$.  (In our convention $|g\rangle$ is assigned energy $- \epsilon / 2$ and $|e\rangle$ is assigned energy $\epsilon / 2$.)  The transition rate from $|e\rangle$ to $|g\rangle$ is $\gamma (\bar{n} + 1)$ and that from $|g\rangle$ to $|e\rangle$ is $\gamma \bar{n}$ where $\bar{n} = [\exp(\beta \epsilon) - 1]^{-1}$ is the bath occupation and $\gamma$ is the zero-temperature decay rate. At high temperatures, $\bar{n}\gg 1$, these two rates are approximately equal to $\Gamma(\beta)\equiv \gamma\bar{n}$.   

A simple irreversible clock using a \textit{single} two-level system can now be defined operationally by assuming that we have a means of making strong continuous (in time) measurements of the energy of the two-level system. Suppose, for example, we find the system in the ground state at time $t=0$, i.e. we find the random telegraph signal is $Z(0)=-1$. The probability that the system continues in this state for a time $t > 0$ is $Pr(Z(t)=-1)=\Gamma(\beta) e^{-\Gamma(\beta) t}$. We can thus use the estimator
\begin{align}
t_{\text{est.}} =\frac{1}{\Gamma(\beta)}
\end{align}
for time elapsed with parameter $\bar{\Gamma}(\beta)$, which we stress is dependent upon temperature.  In the high-temperature limit, this becomes
\begin{equation}
	t_{\text{est.}} = \frac{\epsilon}{\gamma k_BT} .
\end{equation}

In contrast to the example from radiocarbon dating, here the relative error in $t_{\text{est.}}$ is limited by the relative error in $T$.  That is, in order for this system to define an accurate temporal scale, we require a good thermometer.  Conversely, we may use this system as a thermometer where the temperature estimate is
\begin{equation}
	T_{\text{est.}} = \frac{\epsilon }{\gamma k_B t} ,
\end{equation}
so one may equally well say that in order for this system to define an accurate temperature scale, we require a good clock.  In summary,
\begin{equation}
	\frac{\delta t_{\text{est.}}}{t_{\text{est.}}} + \frac{\delta T_{\text{est.}}}{T_{\text{est.}}} = 0 .
\end{equation}

We have assumed that we can continuously monitor the energy of the two-level system  without disturbing local thermodynamic equilibrium. This will require a system-measurement interaction which commutes with the  Hamiltonian of the two-level system---a quantum non demolition interaction. As we will see below, such a measurement does not change the rate equations for the occupation probabilities and thus the steady state achieved by thermal equilibrium is not affected by the measurement.

The most general way to express the limit to the accuracy of a clock of any kind is based on a parameter estimation bound for a density operator of the clock system parameterised by time $\rho(t)$. This is given in terms of the rate of change of statistical distance\cite{BraMil},
\begin{align}
\delta t ^2 \geq \left (\frac{ds}{dt}\right )^{-2}
\end{align}
 where 
 \begin{align}
\frac{ds}{dt}= {\rm tr}\left (\frac{d\rho}{dt}{\cal L}_\rho\left [\frac{d\rho}{dt}\right ]\right )
\end{align}
and where the operator ${\cal L}_\rho$ is defined as the inverse of the operator ${\cal R}_\rho[\hat{A}]=(\rho\hat{A}+\hat{A}\rho)/2$. For a single two-level system, the statistical distance is given by the simple expression
\begin{align}
\left (\frac{ds}{dt}\right )^2=\sum_{\mu=0}^3\left (\frac{dx_\mu}{dt}\right )^2
\end{align}
with the definitions $x_1=\langle\sigma_x\rangle,\ x_2=\langle\sigma_y\rangle,\  x_3=\langle\sigma_z\rangle,\ x_0^2=(1-(x_1^2+x_2^2+x_3^2))$.

In the case of a two level system interacting with a thermal bath, the dynamics is given by the master equation
\begin{align}
\frac{d\rho}{dt}= \gamma(\bar{n}+1){\cal D}[\sigma_-]\rho+\gamma\bar{n}{\cal D}[\sigma_+]\rho
\end{align}
where ${\cal D}[A]\rho = A\rho A^\dagger -\frac{1}{2}(A^\dagger A\rho+\rho A^\dagger A)$ with $\sigma_+=(\sigma_-)^\dagger =|e\rangle\langle g|$.  In this case we find that
\begin{eqnarray}
\dot{x}_1 & = & -\Gamma x/2\\
\dot{x}_2 & = & -\Gamma y/2\\
\dot{x}_3 & = & -\Gamma x_3-\gamma
\end{eqnarray}
with $\Gamma=\gamma(2\bar{n}+1)$. The solutions are 
\begin{eqnarray}
x_1(t) & = & x_1(0)e^{-\Gamma t/2}\\
x_2(t) & = &  x_2(0)e^{-\Gamma t/2}\\
x_3(t) & = & x_{3,\infty}+(x_3(0)-x_{3,\infty})e^{-\Gamma t}
\end{eqnarray}
where $x_{3, \infty}=-1/(2\bar{n}+1)$  is the steady state value of $x_3(t)$.  
For an initial thermal state, $x_1(0)=x_2(0)=0$ and $x_3(0)=-\tanh(\beta\epsilon/2)$. The statistical distance is then given by 
\begin{align}
\label{}
\left (\frac{ds}{dt}\right )^2=\Gamma^2(x_3(0)-x_{3,\infty})^2e^{-2\Gamma t}(1-x_3(t)^2)^{-1}
\end{align}

If all we know is that a two-level system is in thermal equilibrium with a heat bath at temperature $T$, Eq.9 implies that we cannot use it to implement a good clock. Strong continuous measurement of the energy changes this as the conditional state that results will be far from the stationary state of thermal equilibrium. Under strong continuous measurement of the energy, as we assumed above, the conditional value of $x_3(t)^2=1$ as the system makes random telegraph transitions between energy eigenstates. In that case we have the bound $\delta t^2 \geq 0$. The simple protocol described above can achieve this only in the limit of very high temperature.

\section{Quantum thermalisation and Newton's law of cooling}
\label{qubit-cooling}

 Newton's law of cooling describes, under suitable assumptions, how  two materials beginning at different temperatures and placed in thermal contact eventually come to the have the same temperature. It says that the rate of change of temperature for each material is proportional to the temperature difference between them.

\subsection{A two qubit model.}
We now give a quantum model for Newton's law of cooling based on a pair of identical two-level systems---qubits---each initially in thermal equilibrium with two distinct heat baths at temperatures  $T_1 > T_2$.  The input state is $\rho_{\text{in.}} = \rho_{\text{th.}}^1(\beta_1) \otimes \rho_{\text{th.}}^2(\beta_2)$ where $\rho_{\text{th.}}$ is given by Eq.~\eqref{eq:thermalState}.  We suppose that each ensemble is then isolated from their respective thermal environments and allowed to interact with each other.  To begin with, let us describe this interaction by a map from input to output states---we shall extend this description to a continuous time flow later.  That is, we imagine that the pair of two-level systems is injected into a device described by a completely-positive trace-preserving map $\mathcal{D}$ on the tensor-product space of the pair of two-level systems, which produces the output state $\rho_{\text{out.}} = \mathcal{D}(\rho_{\text{in.}})$.  Following Brunner et al.~\cite{Brunner}, we assume that the device $\mathcal{D}$ acts as a non-deterministic swap operation: with probability $\eta$ it implements the transformation $|x\rangle_1\langle x|\otimes |y\rangle_2\langle y|\rightarrow|y\rangle_1\langle y|\otimes |x\rangle_2\langle x|$ where $x, y = e, g$, i.e.,
\begin{equation}
	\mathcal{D}(|x\rangle_1\langle x|\otimes |y\rangle_2\langle y|) = (1-\eta)|x\rangle_1\langle x|\otimes |y\rangle_2\langle y|+\eta |y\rangle_1\langle y|\otimes |x\rangle_2\langle x|
\end{equation}
The input state is thereby mapped to
\begin{equation}
	\rho_{\text{out.}} = (1 - \eta) \rho^{1}_{\text{th.}}(\beta_1) \otimes \rho^{2}_{\text{th.}}(\beta_2) + \eta \rho^{1}_{\text{th.}}(\beta_2) \otimes \rho^{2}_{\text{th.}}(\beta_1) .
\end{equation}
In order for the reduced density matrices of each subsystem after the map to be identical we require $\eta = 1/2$, whereupon $\operatorname{tr}_{1} \rho_{\text{out.}} = \operatorname{tr}_{2} \rho_{\text{out.}} = \rho_{\text{th.}}(\beta_{\text{out.}})$ where
\begin{equation}
	\tanh(\beta_{\text{out.}} \epsilon / 2) = \frac{1}{2} [\tanh(\beta_{1} \epsilon / 2) + \tanh(\beta_{2} \epsilon / 2)] .
\end{equation}
The mixed ensemble has come to a new thermal state with inverse temperature $\beta_{\text{out.}}$.  If the initial temperatures are very high, then $\beta_{\text{out.}} \approx (\beta_1 + \beta_2) / 2$, and if in addition the temperature difference is very small then $T_{\text{out.}} \approx (T_1 + T_2) / 2$.

An alternative way to write down the action of $\mathcal{D}$ is by introducing the unitary swap operator $S$:
\begin{equation}
	S = | e \rangle_1 \langle e | \otimes | e \rangle_2 \langle e | + | e \rangle_1 \langle g | \otimes | g \rangle_2 \langle e | + | g \rangle_1 \langle e | \otimes | e \rangle_2 \langle g | + | g \rangle_1 \langle g | \otimes | g \rangle_2 \langle g | .
\end{equation}
Then, $\mathcal{D}(\rho) = (1 - \eta) \rho + \eta S \rho S$.  Now we may make the process dynamical by supposing that a swap happens at random times with rate $\gamma$, i.e., in an infinitesimal time $dt$ a swap occurs with probability $\gamma dt$, otherwise nothing happens.  The corresponding change in the ensemble is then simply the action of $\mathcal{D}$ with $\eta = \gamma dt$: $\rho(t + dt) = (1 - \gamma dt) \rho(t) + \gamma dt S \rho(t) S$.  This implies $\rho$ obeys the equation of motion
\begin{equation}
	\frac{d\rho}{dt} = \gamma (S \rho S - \rho) ,\label{eq:2LSEoM}
\end{equation}
which has solution
\begin{equation}
	\rho(t) = e^{- \gamma t} [\cosh(\gamma t) \rho(0) + \sinh(\gamma t) S \rho(0) S] .
\end{equation}
Taking $t \rightarrow \infty$ yields the stationary state $\rho(\infty) = [\rho(0) + S \rho(0) S] / 2$, which is simply the action of $\mathcal{D}$ with $\eta = 1/2$, whereupon for $\rho(0) = \rho_{\text{in.}} = \rho_{\text{th.}}^1(\beta_1) \otimes \rho_{\text{th.}}^2(\beta_2)$ we recover $\rho(\infty) = \rho_{\text{out.}} = [\rho_{\text{th.}}^1(\beta_1) \otimes \rho_{\text{th.}}^2(\beta_2) + \rho_{\text{th.}}^1(\beta_2) \otimes \rho_{\text{th.}}^2(\beta_1)]/2$ and the reduced state of each two-level system is a thermal state with inverse temperature $\beta_{\text{out.}}$ as above.  On the other hand, let us consider the average energy of each two-level system $E_i = (\epsilon / 2) \operatorname{tr} (\rho \sigma_z^i)$ where $\sigma_z^i = | e \rangle_i \langle e | - | g \rangle_i \langle g |$ and $i = 1, 2$.  From the master equation~\eqref{eq:2LSEoM}, these obey the equations of motion
\begin{align}
	\frac{d E_1}{dt} &= - \gamma (E_1 - E_2) \text{ and}\label{eq:energyEoM1}\\
    \frac{d E_2}{dt} &= \gamma (E_1 - E_2) .\label{eq:energyEoM2}
\end{align}
Note that the total energy of the two systems is a constant of the motion as required for complete thermal isolation.  We can parameterise the reduced density matrix of each two-level system at all times in terms of a time-dependent inverse temperature $\beta_i$ via $E_i = (\epsilon / 2) \tanh(- \beta_i \epsilon / 2)$ where $i = 1, 2$.  At high temperatures Eqs.~\eqref{eq:energyEoM1} and~\eqref{eq:energyEoM2} then become
\begin{align}
	\frac{dT_1}{dt} &= - \gamma \left( \frac{T_1}{T_2} \right) (T_1 - T_2) \text{ and}\\
    \frac{dT_2}{dt} &= \gamma \left( \frac{T_2}{T_1} \right) (T_1 - T_2) .
\end{align}
Newton's law of cooling thus results only for high temperatures, $T_1, T_2 \gg \epsilon$, and only if the difference in temperature is small, $T_1 \approx T_2$.  The more general expression is given by the rate of change of the average energy difference between the two ensembles and this will form the basis for a Mach clock implementation.

How does one use a model like this to tell time? In the analogy with Mach's clock, the rate of change of the mean energy, Eqns.(\ref{eq:energyEoM1},\ref{eq:energyEoM2}), suggest that we should consider measuring the energy difference between the two ensembles in place of energy measurements. Equivalently, one can monitor the population of the energy eigenstates.   

One approach is similar to the example of radio carbon dating. Consider a large ensemble, $N \gg 1$, of identical pairs of two-level systems each of which is subject to  continuous, very accurate, energy measurements. At the initial time, just as the process is starting, we check to see how many of the pairs are in opposite states, say $|g\rangle_1\otimes|e\rangle_2$. The  number of such systems is approximately $N_{ge}(0)\approx Np_g^{(1)}p_e^{(2)}$. We then simply wait and count how many of these systems have remained in this state for the entire duration, $N_{ge}(t)$. This is determined simply by the probability that all such systems have never experienced a swap operation, which is  $(1-e^{-\gamma t})$. Thus we can use the estimate 
\begin{equation}
t_{est}= \frac{1}{\gamma}\ln \left [1-\frac{N_{ge}(0)}{N_{ge}(t)}\right ]
\end{equation}
The fractional error in this estimate is the same as for the case of any Poisson process
\begin{equation}
\frac{\delta t_{est}}{t_{est}}=\frac{1}{\sqrt{\gamma t_{est}}}
\end{equation}

One may object that the protocol just described assumes the ability to continuously monitor the energy of each system, but we have not taken this process into account explicitly in the dynamics.  It also explicitly refers to an ensemble yet it should be possible to consider a system made up of just two qubits subject to the same continuous irreversible dynamics. 

To address these points we now consider a model in which we include weak continuous measurements of $\sigma_z$ on each qubit.  
The master equation describing the ensemble average dynamics is then given by\cite{HMW_GJM} 
\begin{equation}
\frac{d\rho}{dt} = \gamma(S\rho S-\rho)+\Gamma \sum_{k=1}^2{\cal D}[\sigma_z^{(k)}]\rho
\end{equation}
where $\Gamma$ is the measurement strength parameter. Note that these measurements do not change the dynamics of the 
occupation probabilities of the qubit states as the measurement operator is diagonal in this basis. Thus adding the measurement does not change the ensemble average dynamics for systems that start in product thermal states. 

The observed measurement records $y_k(t)$ are classical stochastic process that obey the Ito stochastic differential equation for the measurement current, 
\begin{equation}
\label{records}
dy_k(t)=z_k(t) dt +\frac{1}{\sqrt{8\Gamma}} dW_k(t)
\end{equation}
where $dW_k$ are independent Wiener increments and $z_k(t)=\langle \sigma_z^{(k)}\rangle_c$ is the conditional mean value of $\sigma_z^{(k)}$ up to time $t$ conditioned on the entire previous history of the measurement record $y_k(t)$. It is given by 
\begin{equation}
\langle \sigma_z^{(k)}\rangle_c ={\rm tr}[ \sigma_z^{(k)}\rho_c(t)]
\end{equation}
where $\rho_c(t)$, the conditional state, satisfies the stochastic master equation
\begin{equation}
d\rho = \gamma(S\rho S-\rho)dt +\Gamma \sum_{k=1}^2{\cal D}[\sigma_z^{(k)}]\rho \ dt+\sqrt{\Gamma}\sum_{k=1}^2{\cal H}[\sigma_z^{(k)}]\rho\  dW_k(t)
\end{equation}

The stochastic terms, that represent the conditional evolution, make this a nonlinear dynamical system. The effect of the stochastic terms is to drive the system towards the energy eigenstates. To see this note that when the system enters one of the four energy eigenstates, the noise term vanishes.  

The measurement record has two sources of stochasticity; (i) the added white noise due to the measurement that scales as $\Gamma^{-1/2}$ and (ii), due to the stochastic dynamics of $z_k(t)$ as we describe below.  In analogy with the Mach clock we will consider the difference measurement current defined by  
\begin{eqnarray}
 \frac{dy_-}{dt} & = & \frac{dy_1}{dt}-\frac{dy_2}{dt}\\
  & = & (z_1(t)-z_2(t))  +\frac{1}{\sqrt{8\Gamma}} \xi(t)
 \end{eqnarray}

We first find $z_-(t)=z_1(t)-z_2(t)$. Using the stochastic equations for the 
occupation probabilities we find that
\begin{eqnarray}
dz_1(t) & = & -\gamma(z_1-z_2)dt+2\sqrt{\Gamma} dW_1(t) (1-z_1^2)\\
dz_2(t) & = & \gamma(z_1-z_2)dt+2\sqrt{\Gamma} dW_2(t) (1-z_2^2)
\end{eqnarray}
Thus,
\begin{equation}
dz_-(t)=-2\gamma z_-(t)dt+2\sqrt{\Gamma}dW_1(t)(1-z_1^2)-2\sqrt{\Gamma} dW_2(t)(1-z_2^2)
\end{equation}
Note that if we now  average over the noise, we reproduce the ensemble average equations of motion  in Eqs.(\ref{eq:energyEoM1},\ref{eq:energyEoM2}).

 Ignoring the noise,  the systematic dynamics is linear and $z_1(t)+z_2(t)$ is a constant of the motion
 and 
 \begin{equation}
 z_1(t)-z_2(t)= (z_1(0)-z_2(0))e^{-2\gamma t}
 \end{equation}
 and the systematic part of the measurement current $ \frac{dy_-}{dt}$ simply decays to zero. Due to the noise, the ability to distinguish the two measurement currents becomes increasingly difficult. In fact, if the initial temperatures are not sufficiently different, we would not be able to distinguish the two measurement currents even at short times due to the white noise added by the measurement itself. 

\begin{figure}[htbp] 
   \centering
   \includegraphics[scale=0.5]{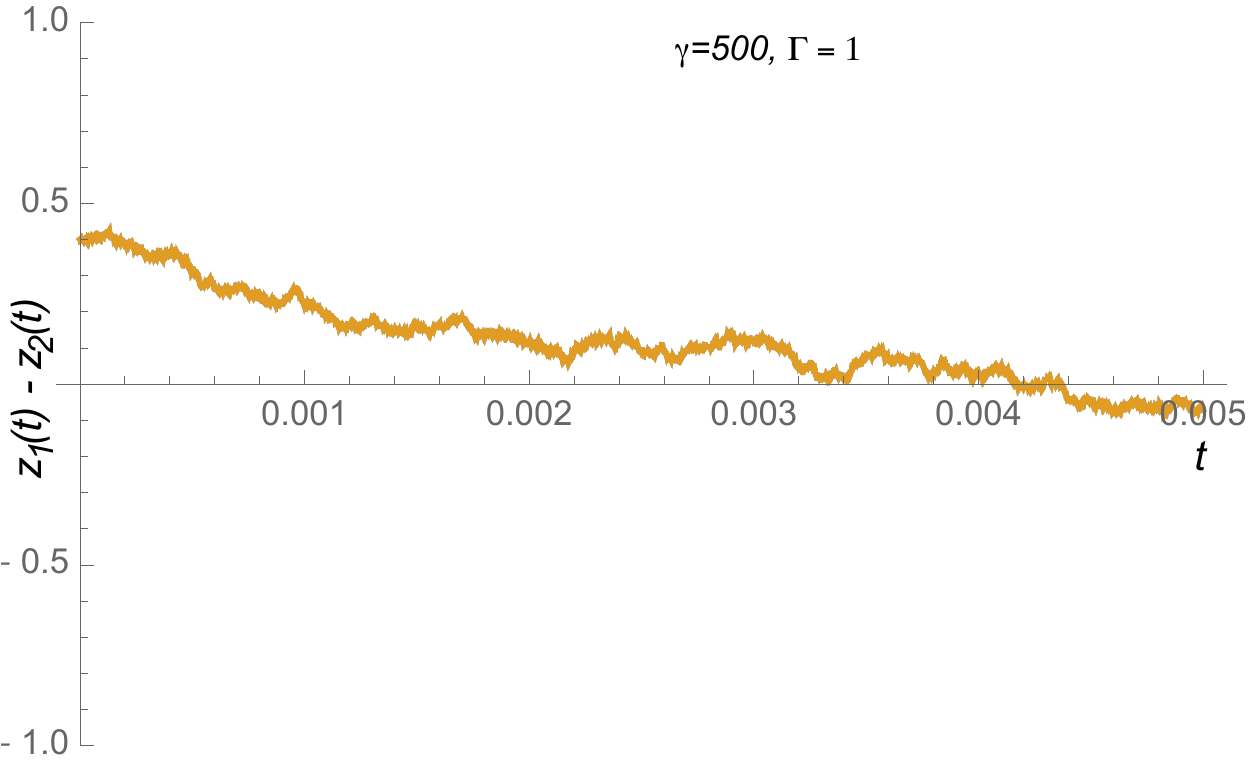} 
   \caption{A simulated conditonal average of the difference $z_-(t)$ in the limit of a slow weak continuous measurement. $\gamma=500,\ \ \Gamma=1$ }
   \label{fig:weak-meas}
\end{figure}

 We will first consider the case in which the measurement rate is much smaller than the swap rate,  $\Gamma \ll \gamma$.  A typical example is shown in Fig. \ref{fig:weak-meas}.
  Over a time $t$ such that $\gamma t \ll 1$,  the solution can be approximated by
 \begin{eqnarray}
z_1(t) & = & z_1(0)-(z_1(0)-z_2(0))\gamma t+2\sqrt{\Gamma}\delta W_1(t) (1-(z_1(0))^2)\\
z_2(t) & = & z_2(0)+(z_1(0)-z_2(0))\tilde{\gamma}\tau+2\sqrt{\Gamma}\delta W_1(t) (1-(z_2(0))^2)
\end{eqnarray}
where we have approximated the  noise as a constant determined by the mean and variance of the independent Gaussian random variables $\delta W_k(t)$ with zero mean and variance, $Var(\delta W_k(t)=t$ and, by assumption $\Gamma t\ll 1$. The initial conditional averages are simply given by the initial thermal product states so that $z_j(0)=\tanh (\beta_j\epsilon/2)$.

Define the random variable
\begin{equation}
S(t)=\frac{z_1(t)-z_2(t)}{z_1(0)-z_2(0)}
\end{equation}
The ensemble average of this quantity is 
\begin{equation} 
\overline{S(t)}=1-2\gamma t
\end{equation}
However for short times the fluctuations in $S(t)$are given by
the standard deviation of the gaussian random variable
$2\sqrt{\Gamma}\delta W_1(t) (1-(z_1(0))^2)-2\sqrt{\Gamma}\delta W_1(t) (1-(z_2(0))^2)$.  
This is 
\begin{equation}
\Delta S(t) = 2 \sqrt{\mu\Gamma t}
\end{equation}
where 
\begin{equation} 
\mu= \frac{(1-(z_1(0))^2)^2+(1-(z_2(0))^2)^2}{(z_1(0)-z_2(0))^2}
\end{equation}

At high temperatures, the initial states are close to the identity and $z_j(0)\approx \beta_j\epsilon/2$. In that case
\begin{equation}
\mu=2\left [\frac{\epsilon}{2k_B}\left (\frac{1}{T_2}-\frac{1}{T_1}\right )\right ]^{-2}
\end{equation}
This quantity has an interesting interpretation in terms of the Kullback-Leibler divergence\cite{Cover}, an information-theoretical measure of the statistical distinguishability of two thermal states at different temperatures.  Define
\begin{equation}
D=p_g^{(1)}\ln\left (\frac{p_g^{(1))}}{p_g^{(2)}}\right ) + p_e^{(1)}\ln\left (\frac{p_e^{(1))}}{p_e^{(2)}}\right )\ .
\end{equation}
which, for high temperatures reduces to 
\begin{equation}
D\approx \frac{\epsilon}{2k_B}\left (\frac{1}{T_2}-\frac{1}{T_1}\right )
\end{equation}
At high temperatures we thus see that 
\begin{equation}
\Delta S(t) =\frac{\sqrt{8\Gamma t}}{D}
\end{equation}
To make the error in the estimator as small as possible we need to ensure $D \gg \sqrt{2\Gamma t}$. The error will be small in the case that the two ensembles are strongly distinguishable. A good estimate of the time will require $2\gamma t \gg \Delta S(t)$ which is equivalent to the condition
\begin{equation}
\sqrt{2\Gamma/\gamma} \gg D
\end{equation}


Now turn to what we might do in a particular trial of the experiment. 
Suppose we measure a value for $S(t)$. 
We  define the estimator for the elapsed time $\tau$, given a particular measurement record as 
\begin{equation}
t_{est} = (1-S(t))/\gamma
\end{equation}
The noise in this estimate is  given by 
\begin{equation}
\Delta t_{est}  = \Delta S(t)/ \gamma
\end{equation}
where the uncertainty $\Delta S(t)$ is  given above. 

 There is another important limit corresponding to``quantum jumps". This occurs when the `measurement rate' is much faster than the rate of probabilistic swaps, $\Gamma \gg \gamma$ so that $\gamma/\Gamma\rightarrow 0$.  In this limit, the measurement rapidly localises $z_k$ on $\pm 1$ with the relative fraction ending up on $z_k= 1$ given by the probability $p^{(k)}_{e}(0)$ and the fraction ending up on $-1$ given by $p^{(k)}_{g}(0)$.   This is the limit of  projective measurements. If each system initially localises on a different $\sigma_z$ eigenstate we simply see Poisson distributed swapping, otherwise there is no dynamics at all. In this limit we are back to the Poisson distributed jumps method for estimating elapsed time like radiocarbon dating and thermalisation of a two-level system. 

\subsection{An optomechanical model.} 
\begin{figure}[htbp] 
   \centering
   \includegraphics[scale=0.7]{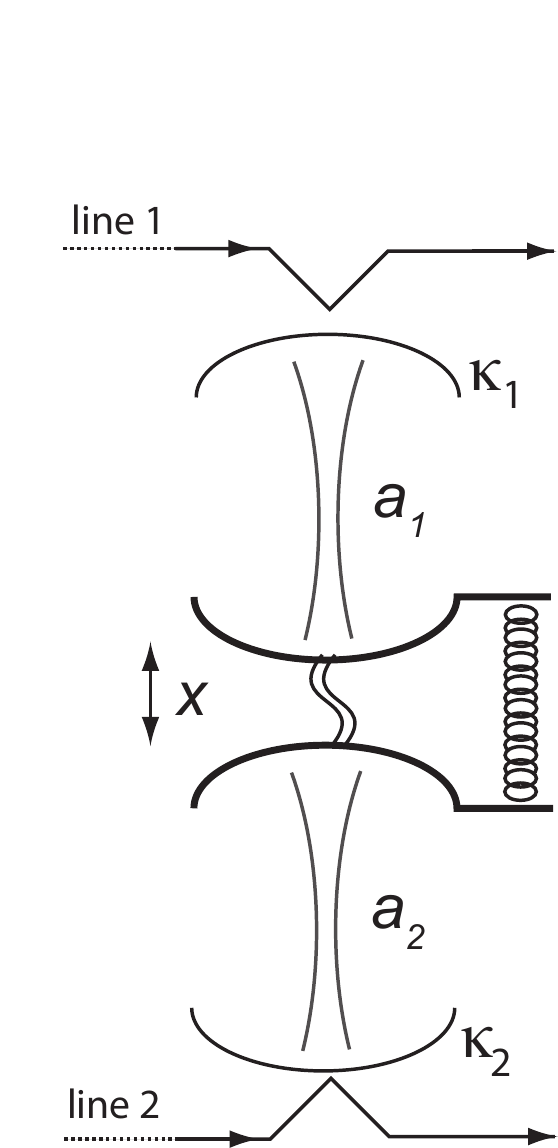} 
   \caption{Two optical cavies are coupled in such a way that the coupling rate is modulated by the displacement of a mechanical 
   resonator. The cavity photon decay rates are $\kappa_1,\kappa_2$ and $x$ represents a small displacement in the distance between the cavities while line 1 and line 2 refer to waveguides coupled to cavity 1 and 2 respectively. The operators $a_1,a_2$ are bosonic annihilation operators for photons in each cavity.     }
   \label{fig1}
\end{figure}

We now turn to a more experimentally accessible model to illustrate a cooling law in a quantum system. In Chang et al \cite{Chang} a model for two coupled optical cavities with a  coupling mediated by a mechanical system was presented. The scheme is depicted in Fig. \ref{fig1}.
The Hamiltonian for this systems is given by
\begin{equation}
\label{full-ham}
H=\hbar\omega_1 a_1^\dagger a_1+\hbar\omega_2 a_2^\dagger a_2+\hbar\omega_m b^\dagger b +\hbar g \hat{X}(a_1^\dagger a_2+a_1 a_2^\dagger)
\end{equation}
where $a_k, a_k^\dagger$ are the annihilation and creation operators for the cavity fields, $b,b^\dagger$ are the annihilation and creation operators for the mechanical resonator, $\hat{X}= b+b^\dagger$, $\omega_k$ are the resonant frequencies of each cavity mode and $\omega_m$ is the resonate frequency of the mechanical element and $g$ is the coupling constant. The interaction Hamiltonian in Eq. (\ref{full-ham}) plays the role of the non-deterministic swap operation in the abstract qubit model of the previous section. A similar interaction was used by Levy et al. \cite{Levy} in their model of a quantum refrigerator. 

We now move to an interaction picture at the optical and mechanical frequencies,
\begin{equation}
H_I(t)=\hbar g \hat{X}(t)(a_1^\dagger a_2e^{i\Delta t}+a_1 a_2^\dagger e^{-i\Delta t})
\end{equation}
where $\hat{X}(t) = b e^{-i\omega_m t}+b^\dagger e^{i\omega_m t}$ and $\Delta =\omega_1-\omega_2$. There are two choices for resonant interactions, $\Delta =\pm\omega_m$ for which we can make the rotating wave approximation to obtain
\begin{eqnarray}
\label{rwa-om-ham}
H_+ & = & \hbar g(b a_1^\dagger a_2+b^\dagger a_1 a_2^\dagger)\\
H_- & = & \hbar g (b^\dagger a_1^\dagger a_2 + ba_1 a_2^\dagger)
\end{eqnarray}
These Hamiltonians describe Raman transitions in which a photon is moved from cavity $2$ to cavity $1$ by {\em absorbing} a mechanical excitation in the case of $H_+$ and {\em emitting} a mechanical excitation in the case of $H_-$. Energy is conserved by the mechanical system exchanging excitations with its heat bath. Note that exchanging the labels for optical cavities makes these Hamiltonians equivalent. We will see below that the two choices for  the detuning correspond to opposite directions of heat flow.

In reality both the optical cavities and the mechanical resonator are damped. In the case of the mechanical resonator it is also coupled to a non-zero temperature that bath. These may be described using the master equation
\begin{equation}
\frac{d\rho}{dt}=-\frac{i}{\hbar}[H,\rho]+\kappa_1{\cal D}[a_1]\rho+\kappa_2{\cal D}[a_2]\rho+\gamma(\bar{n}+1){\cal D}[b]\rho+\gamma \bar{n}{\cal D}[b^\dagger]\rho
\end{equation}
where $\hat{H}$ is given by either of the Eqs. (\ref{rwa-om-ham}).  In seeking a quantum model analogous to the Mach clock, we are primarily interested in how the cavities evolve when they are coupled through the opto mechanical interaction. We will thus begin by setting the cavity decay rates to zero. 

 If the mechanics is coupled to a high temperature heat bath, it is rapidly thermalised to a steady state that is little effected by the coupling to the optical cavities.  To be more precise, if $\gamma \bar{n} \gg g$ where $\gamma, \bar{n}$ are the mechanical damping rates and the thermal occupation of the mechanical bath respectively, we can adiabatically eliminate the mechanical degree of freedom to obtain an effective master equations for the cavity fields alone (see appendix)
\begin{equation}
\label{adiabatic-me}
\frac{d\rho}{dt} = {\cal L}_\pm\rho
\end{equation}
where the two Lindbald super-operators corresponding to the interaction Hamiltonians $H_\pm$ are
\begin{eqnarray}
{\cal L}_+\rho & = & \Gamma(\bar{n}+1){\cal D}[a_1 a_2^\dagger]\rho+ \Gamma\bar{n}{\cal D}[a_1^\dagger a_2]\rho\\
{\cal L}_- \rho & = & \Gamma(\bar{n}+1){\cal D}[a_1^\dagger a_2]\rho+ \Gamma\bar{n}{\cal D}[a_1 a_2^\dagger]\rho
\end{eqnarray}
with ${\cal D}[A]\rho \equiv A\rho A^\dagger -A^\dagger A\rho/2 -\rho A^\dagger A/2$ and 
\begin{equation}
\Gamma= \frac{4g^2}{\gamma}
\end{equation}
Eq. (\ref{adiabatic-me}) is equivalent to the model in Levy et al.\cite{Levy}, Eq. (8)) with the role of hot and cold baths interchanged when we interchange ${\cal L}_+$ and ${\cal L}_-$.

We will restrict the discussion in the paper to the case of ${\cal L}_+$. The master equation in Eq. (\ref{adiabatic-me}) describes two conditional Poisson processes, $dN_{12}$ and $dN_{21}$. The jump operators\cite{WisMil} for these processes are
\begin{eqnarray}
{\cal J}_{12}\rho & = &     a_2^\dagger a_1\rho a_2 a_1^\dagger\\
{\cal J}_{21}\rho & = &  a_1^\dagger a_2\rho a_1 a_2^\dagger
\end{eqnarray}
In the first of these, a photonic excitation is removed from cavity $1$ and added to cavity $2$, in the second, the opposite occurs. The rates for the corresponding Poisson process are given by 
\begin{eqnarray}
{\cal E}(dN_{12}) & = &   \Gamma(\bar{n}+1)\ {\rm tr}[{\cal J}_{12}\rho]dt=\Gamma(\bar{n}+1)\langle \hat{n}_1(\hat{n}_2+1)\rangle dt\\
{\cal E}(dN_{21}) & = & \Gamma\bar{n}\ {\rm tr}[{\cal J}_{21}\rho]dt=\Gamma\bar{n}\langle \hat{n}_2(\hat{n}_1+1)\rangle dt 
\end{eqnarray}
We can see that the ratio
\begin{equation}
\label{Gibbs-factor}
r=\frac{\bar{n}}{\bar{n}+1}=e^{-\beta \hbar\omega_m}=e^{-\beta \hbar(\omega_1-\omega_2)}
\end{equation}
is the thermal Boltzmann factor with $\beta$ the temperature of the mechanical system at frequency $\omega_m$ and the last expression results from noting that we are  considering the case for which  $\omega_1-\omega_2=\omega_m$. In Levy et al.\cite{Levy} this would correspond to regarding cavity $1$ as the hot system and cavity $2$ as the cold system. We will use a similar definition in what follows and assume that initially the cavities are prepared in thermal states with $T_1>T_2$.  Thus the Poisson process $dN_{12}$ descries a transition from hot to cold while $dN_{21}$ describes a transition for cold to hot. 

We can define a net number current (or flux) by  the classical stochastic process
\begin{equation} 
i(t)dt = dN_{12}-dN_{21}
\end{equation}
The change in the energy of the cavities in this time is
\begin{equation} 
dE_c(t) =-\hbar\omega_m(dN_{12}-dN_{21})
\end{equation}
The ensemble average of this current is given by 
\begin{eqnarray} 
\label{average-current}
\overline{i(t)}  & = &  \Gamma(\bar{n}+1)\langle \hat{n}_1(\hat{n}_2+1)\rangle-\Gamma\bar{n}\langle \hat{n}_2(\hat{n}_1+1)\rangle\\
& = & \Gamma(\bar{n}+1)\langle \hat{n}_1\rangle -\Gamma\bar{n}\langle \hat{n}_2\rangle+\Gamma\langle \hat{n}_1\hat{n}_2\rangle
\end{eqnarray}
On the other hand the change in the energy of the mechanics over this time interval is 
\begin{equation} 
dE_m(t) =\hbar\omega_m(dN_{12}-dN_{21})
\end{equation}
as a transfer of a photon from cavity $2$ to $1$ requires an absorption of one excitation from the mechanics, and  a transfer of a photon from cavity $1$ to $2$ requires the emission of one excitation into the the mechanics. The average current then determines the rate at which phonons enter the mechanics.  It is clear that $dE_c(t)+dE_m(t)=0$ for conservation of energy. Thus $\overline{i(t)}$ is the negative of the rate of change of energy in the optical  system.   

\begin{figure}[htbp] 
   \centering
   \includegraphics[scale=1.0]{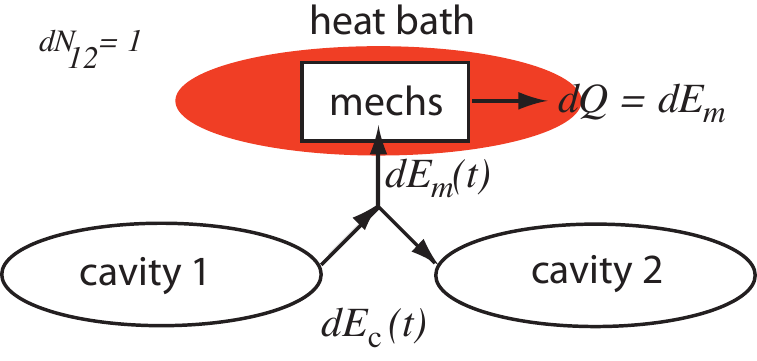} 
   \caption{A schematic representation of the energy exchanges involved when one photon is transferred from cavity-1 to cavity-2, i.e. $dN_{12}=1$. This requires the emission of a single phonon into the mechanical resonator. To maintain the resonator in thermal equilibrium, a small exchange of heat energy must take place between the mechanical degree of freedom and the heat bath with which it is in contact. }
   \label{fig2}
\end{figure}

The mechanical degree of freedom  is held in contact with a heat bath. If a single phonon enters/exits the mechanical system it is no longer in thermal equilibrium but is rapidly restored to thermal equilibrium by the exchange of heat with the thermal bath with which it is in contact. In this way a classical stochastic heat transfer $dQ$ is conditioned on elementary quantum tunnelling events. We illustrate this relation in Fig. \ref{fig2} for the case of a transfer of a single photon from cavity-1 to cavity-2 corresponding to the event $dN_{12}=1$. 

The nonlinear dependence on photon number in the expressions for the rates of each stochastic process is indicative of nonlinear cooling/heating. This can be checked by computing the rate of change of average photon number in each cavity using the master equation in Eq. (\ref{adiabatic-me}). 
 For example the rate of change of photon number difference is 
\begin{equation}
\label{number-difference}
\frac{d\langle \hat{n}_2-\hat{n}_1\rangle }{dt}=2\Gamma\langle \hat{n}_2\hat{n}_1\rangle+2\Gamma(\bar{n}+1)\langle\hat{n}_1\rangle-2\Gamma \bar{n}\langle \hat{n}_2\rangle
\end{equation}
While $\hat{n}_2+\hat{n}_1$ is a constant of motion.   Comparing these results to  Eq. (\ref{average-current}) we see that the ensemble average of the current is 
\begin{equation}
\overline{i(t)} = \frac{1}{2}\frac{d\langle \hat{n}_2-\hat{n}_1\rangle}{dt}
\end{equation}

\section{The Optomechanical Mach clock. }
In order to turn this model into a Mach clock we need to introduce an analogue to a temperature measurement on each of the cavity modes. The temperature is a thermodynamic quantity for macroscopic systems and ensemble averages of microscopic systems. In our model the temperature of the mechanics is well defined as it is in thermal equilibrium with a heat bath.  How to generalise thermodynamics quantities to the case of a single quantum system not in thermal equilibrium is one of the key questions in the field of quantum thermodynamics\cite{Jarzynski}.    As for the qubit model, we consider continuous quantum non demotion measurements of photon number in each cavity as a quantum equivalent to temperature measurements in the macroscopic thermodynamic case.   Such measurements, when averaged over all results, do not change the photon number statistics in each cavity.  

Returning to Eq. (\ref{average-current}) and Eq. (\ref{number-difference})  we see that we need to make a continuous time measurement of the photon number difference between the two cavities. With this in mind we define the operators
\begin{eqnarray}
 \hat{J}_+   & = & \hat{J}_-^\dagger =a_1^\dagger a_2\\
 \hat{J}_z  & = & \frac{1}{2}(\hat{n}_1-\hat{n}_2)
%
 \end{eqnarray}
 These operators have the same algebra as $su(2)$ with the Casimir invariant
 \begin{equation}
 \hat{J}^2 =\frac{\hat{N}}{2}(\frac{\hat{N}}{2}+1)
 \end{equation}
 where $\hat{N}= \hat{n}_2+\hat{n}_1$ is the total photon number operator. 
 
 In terms of the su(2) generators the master equation for the case $\omega_1-\omega_2=\omega_m$ in Eq. \ref{adiabatic-me} can be written
 \begin{equation}
 \label{pseudo-spin}
 \frac{d\rho}{dt} = \Gamma (\bar{n}+1){\cal D}[\hat{J}_-]\rho+\Gamma \bar{n}{\cal D}[\hat{J}_+]\rho
 \end{equation}
 This equation is structured so that the stochastic dynamics takes place entirely in the subspaces defined by the eigenstates of $\hat{N}$. In other words the total `angular momentum' quantum number, $j=N/2$, is conserved. This is 
 direct consequence of the conservation of $a_1^\dagger a_1+a_2^\dagger a_2$. In this 
 situation, the model in Eq. (\ref{pseudo-spin}) is familiar from quantum optics where is 
 it known as the incoherently driven Dicke model for which an exact steady state is known\cite{Hassan}. 
 
 In the optomechanical model,the optical system starts in a product thermal state
 \begin{equation}
 \rho(0)=(1-\lambda_1)(1-\lambda_2)\sum_{n,m=0}^\infty\lambda_1^n\lambda_2^m|n\rangle_1\langle n|\otimes|m\rangle_2\langle m|
 \end{equation}
 with $\lambda_i=e^{-\beta_i\hbar\omega_i}$, 
 and is thus an incoherent mixture of eigenstates of $\hat{J}^2$. We write the initial product thermal state in terms of the eigenstates of $\hat{J}^2, \hat{J}_z$ as 
\begin{equation}
 \label{initial-state}
 \rho_{12}(0)=\sum_{N=0}^\infty p(N)\sum_{n=0}^Np(n|N)\ |N/2, n-N/2\rangle\langle N/2, n-N/2|
 \end{equation}
 where 
 \begin{eqnarray}
 p(n|N) & = & \frac{(1-\mu)\mu^{n-N/2}}{\mu^{-N/2}-\mu^{N/2+1}}\\
 p(N) & = &\frac{(1-\lambda_1)(1-\lambda-2)}{1-\mu}\left (\lambda_1^N-\mu \lambda_2^N\right )
 \end{eqnarray}
 where $\mu =\lambda_1/\lambda_2$. In these expressions, $N=2j$.

The initial value of $\langle \hat{J}_z(0)\rangle$ is 
 \begin{equation}
 \langle \hat{J}_z(0)\rangle=\frac{1}{2}(\langle \hat{N}_1\rangle-\langle \hat{N}_2\rangle )
 \end{equation}
  where $\hat{N}_i=a_i^\dagger a_i$.
 For high temperatures this is approximately 
 \begin{equation}
  \langle \hat{J}_z(0)\rangle\approx\frac{1}{2}\left (\frac{k_BT_1}{\hbar\omega_1}-\frac{k_BT_2}{\hbar\omega_2}\right )
  \end{equation}
  Typically this will be small. 
  
 
From the master equation Eq. (\ref{pseudo-spin}), we find the equation of motion for the photon number difference in the form, 
 \begin{equation}
 \label{su2-me}
 \frac{d\langle \hat{J}_z\rangle }{dt}=-\frac{\Gamma}{4} \langle \hat{N}(\hat{N}+2)\rangle +\Gamma \langle \hat{J}_z^2\rangle-\Gamma (2\bar{n}+1)\langle \hat{J}_z\rangle 
 \end{equation}

 This is certainly not a simple exponential decay. Indeed, for the case of $\bar{n}=0$, and an initial eigenstate of total photon number ($\langle \Delta \hat{N}^2\rangle  =0$), this  is the old problem of superradiance which has a particular kind of non exponential decay, depending on the initial condition\cite{Bonifacio,Braun}. The dynamics in the case of $\bar{n}\neq 0$, with $\langle \Delta \hat{N}^2\rangle  =0$,  has been treated by Hassan et al. \cite{Hassan}. For the case of initial thermal states, we find that,  
 \begin{equation}
 \langle \hat{N}_i^2\rangle =\bar{N}_i+2\bar{N}_i^2
 \end{equation}
 where $\hat{N}_i=a_i^\dagger a_i$.
 Noting that $\hat{N}$ is a constant of motion, we can write 
 \begin{equation}\langle \hat{N}^2\rangle=2(\bar{N}_1^2+\bar{N}_2^2+\bar{N}_1\bar{N}_2)+\bar{N}\ .
 \end{equation}
 In the case that $\bar{N}_1= \bar{N}_2+\epsilon$ with $\epsilon \ll \bar{N}$ we see that 
 \begin{equation}
 \langle \hat{N}^2\rangle/\bar{N}\approx 3\bar{N}/2+1
 \end{equation}
 
To reach a Mach clock limit we need to consider the relationship of $\bar{n}$ and $\bar{N}$ and an appropriate semiclassical limit.  
 It will be convenient to scale this variable in terms of $j=\bar{N}/2$.  
 \begin{equation}
 z=\frac{\langle \hat{J}_z\rangle }{j}
 \end{equation}
 In the limit of $j\gg 1$, $\langle \hat{J}_z^2\rangle/j^2 \approx \langle \hat{J}_z\rangle^2/j^2 =z^2$.  We then find that the semiclassical dynamics is given by 
 \begin{equation}
 \dot{z} = -\frac{3\Gamma}{2}(\bar{N}+1)+ \frac{\Gamma\bar{N}}{2} z^2 -\Gamma(2\bar{n}+1) z
 \end{equation}
 If we now assume that $\bar{n} \gg \bar{N}$  and further that $\bar{n}\gg 1$, we can use the approximation
  \begin{equation}
 \dot{z} \approx  -2\Gamma \bar{n} z
 \end{equation}
which does imply  an exponential decay of the average photon number difference between the two cavities. A similar result was first noted by Hassan et al. \cite{Hassan}. This is the limit in which the optomechanical system obeys a Newton-like cooling law and can function as a Mach clock.  



The average decay of photon number difference gives us an idea of the appropriate limit for this model to function as a Mach clock however it does not describe how we use a single realisation of this system. To answer this we need to consider a weak continuous measurement of $\hat{J}_z$ (see Appendix B).

\section{Discussion and conclusion.}
A clock, be it periodic or non periodic, is a machine and, like all machines, subject to the laws of thermodynamics. Mach's thermal clock is a non periodic clock that operates by the laws of thermodynamics. As Mach emphasised, the time it measures is just as good for kinematics as that measured by periodic mechanical clocks. This equivalence is ultimately due to the implications of the second law of thermodynamics for irreversible dynamical systems.

It might be useful here to consider an explicit example of a periodic irreversible clock. The coherent output of a single mode laser is such an example and forms the essential component of atomic clocks. The key point is that, above threshold, the quadrature phase amplitudes of the laser field relax onto a limit cycle. The steady state distribution of these amplitudes are described by a stationary distribution localised on this limit cycle. On the other hand a continuous monitoring of the field amplitudes at the laser frequency (via heterodyne detection say) shows a slow phase diffusion around the limit cycle. In the ensemble average, the statistics of the diffusing trajectories is described by the stationary distribution function on the limit cycle.  The laser, like the optomechanical Mach clock, is an irreversible system and the phase diffusion on the limit cycle is a reflection of the fluctuation-dissipation theorem and, ultimately, the second law of thermodynamics.

Mach's non periodic clock, like radiocarbon dating, works by using special non equlibrium initial conditions. For thermodynamic systems, this requires implementing physico-chemical processes to drive the state away from equlibrium. In contrast, in this paper we have emphasised that the act of observation itself suffices to prepare a non equlibrium state from a stationary equlibrium state. The future behaviour of the clock is then simply a direct result of the second law of thermodynamics. There is an essential role for the observer in defining a thermal clock (in fact any clock) as well as the special cosmological initial conditions that are conjectured to underlie the second law. 

In the quantum case the role of measurement is inescapable. Rather than measuring a macroscopic thermodynamic quantity (temperature), in our quantum examples we focused on the need to make continuous measurements of energy. Quantum correlations, for example in the Dicke model, lead to departures from the prototypical Mach clock, yet quantum Mach clocks are nonetheless good non-periodic clocks that depend on the second law of thermodynamics for their operation.  

\newpage
\section*{Appendix A}
In this appendix we derive the effective master equations for the optical system by adiabatically eliminating the mechanical degrees of freedom. The key idea is that the mechanical resonator relaxes to a thermal state much faster than the dynamics determined by the interaction with the optical field modes. We will give the example for $H_+$ the derivation for $H_-$ is very much the same.

We begin with the optomechanical interaction written as a time dependent function
\begin{equation}
H_+(t) =\hbar g(b(t)J_+ + b^\dagger(t)J_-)
\end{equation}
where we have defined $J_+=a_1^\dagger a_2 =(J_-)^\dagger$
We will assume that the quantum stochastic processes defined by $b(t)$ and $b^\dagger(t)$ can be 
approximated by the  stationary two-time correlation functions for the annihilation and creation operators of 
the mechanical system subject only to thermalisation. In the long time limit (ignoring initial transients)
we find that
\begin{equation}
b(t) = \sqrt{\gamma} e^{-\gamma t/2}\int_0^t dt'\ e^{\gamma t'/2}b_{in}(t')
\end{equation}
where $b_{in}(t)$ is a quantum white noise process \cite{WisMil}.  The we find for example,
 in the limit that $t\rightarrow \infty$, by the quantum regression theorem
\begin{eqnarray}
\langle b(t) \rangle & = & \langle b^\dagger(t) \rangle =0\\
\langle b^\dagger(t+\tau) b(t) \rangle & = & \bar{n} e^{-\gamma\tau/2}\\
\langle b(t+\tau) b^\dagger(t) \rangle & = & (\bar{n}+1) e^{-\gamma\tau/2}
\end{eqnarray}

The derivation parallels the standard derivation for the master equation, see for example \cite{WisMil}.
To second order in the opto mechanical coupling rate, $g$, we find that the state of the total system satisfies
\begin{eqnarray*}
\frac{d \rho_T}{dt} & = & -ig[b(t)J_+ + b^\dagger(t)J_-,\  \rho(t)\rho_b]\\
&& -g^2\int_0^t dt_1 [b(t)J_+ + b^\dagger(t)J_-, [b(t_1)J_+ + b^\dagger(t_1)J_-,\  \rho_{12}(t)\rho_b]]
\end{eqnarray*}
where we have assumed that we can factorise the total state as $\rho_T(t)\approx \rho(t)\rho_b$ where the mechanical state is thermal.

Taking the partial trace of both sides over the mechanical Hilbert space we see that we find
 \begin{eqnarray*}
\frac{d\rho}{dt} & = &  2g^2 (A {\cal D}[J_-]\rho +B {\cal  D}[J_+]\rho)
\end{eqnarray*}
where,  
\begin{eqnarray*}
A & = &  \int_0^t dt_1\  \langle b(t) b^\dagger(t_1) \rangle\\
B & = &  \int_0^t dt_1\  \langle b^\dagger(t) b(t_1) \rangle
\end{eqnarray*}
Assuming that for $\gamma t\gg 1$ these become
 \begin{eqnarray*}
\frac{d\rho}{dt} & = &  \Gamma(\bar{n}+1) {\cal D}[J_-]\rho +\Gamma\bar{n} {\cal  D}[J_+]\rho)
\end{eqnarray*}

\newpage
\section*{Appendix B}

\begin{figure}[htbp] 
   \centering
   \includegraphics[scale=0.7]{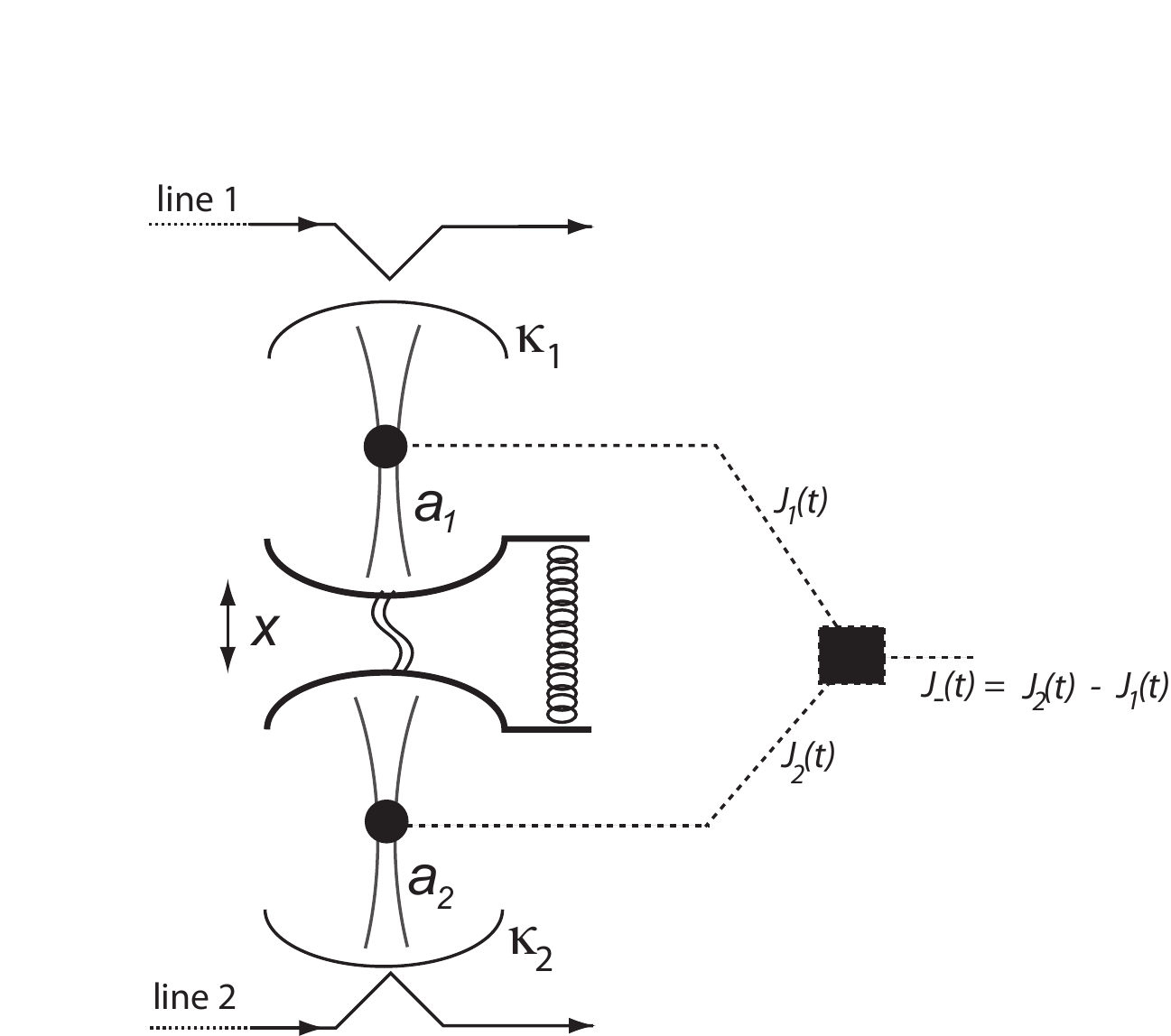} 
   \caption{Photon number measurements are added to each cavity and the difference photon number is continuously measured.  }
   \label{fig3}
\end{figure}

In this appendix we provide an explicit model for the continuous measurement of $\hat{J}_z$. A single two-level atom, driven by a classical field,  interacts dispersively with the field in each cavity.  The raditation emitted from each two-level system is then detected using homodyne detection.  We arrange for the actual readout to correspond to a homodyne detection difference current from the scheme is depicted in Fig. \ref{fig3}. 

We will model the measurement of photon number by introducing  a single qubit into cavity-1 with an interaction Hamiltonian that commutes with the photon number in that cavity. Thus for cavity-j, the Hamiltonian describing the i traction with the measurement system is, 
\begin{equation}
\label{meas-hamiltonian}
H_m=\frac{\hbar\Omega}{2}\sigma_z+\hbar E_0\cos(\omega_0 t)\sigma_x+\kappa a_j^\dagger a_j \sigma_z
\end{equation}
where $\sigma_z =|e\rangle\langle e|-|g\rangle \langle g| $  is the Pauli z-operator $\sigma_x= |e\rangle \langle g|+|g\rangle_ \langle e|$ is the Pauli x-operator, and $\hbar\Omega$ is the free energy difference between the ground state $|g\rangle$ and the excited state $|e\rangle$.   Note that if the cavity field is in thermal equilibrium with a heat bath, this interaction Hamiltonian does not change that as it commutes with the energy operator for the cavity field.  This is the key feature that we need to implement if we are to obtain an appropriate generalisation of a thermometer for a single quantum system. 

The interaction term could model the dispersive interaction between an electric dipole and the cavity field\cite{dispersive}. In terms of an effective spin dynamics, this interaction describes the spin precession of a spin-half dipole around an effective magnetic field proportional to the photon number in the cavity, so the precession frequency will be proportional to the photon number. To measure photon number we need to transduce this frequency and for this we need a  good clock. We are thus led to add a transverse oscillating driving field at frequency $\omega$ and Rabi frequency $E_0$. 

 We now move to an interaction picture at the driving frequency $\omega_0$ for the measurement Hamiltonian to obtain (with the rotating wave approximation)
 \begin{equation}
\label{meas-ham}
H_{I,m}=\frac{\hbar\delta}{2}\sigma_z+\frac{\hbar E_0}{2}\sigma_x+\kappa_j a_j^\dagger a_j \sigma_z
\end{equation}
with $\delta=\Omega-\omega_0$ is the detuning between the bare qubit frequency and the driving field frequency. 
In what follows we will assume that the driving field is resonant with the quit and set $\delta=0$. 

To obtain a signal from the qubit we assume it radiates into an output transmission line at zero temperature so that the full dynamics of the qubit is described by a master equation of the form
\begin{equation}
\frac{d\rho_q}{dt}=-\frac{i}{\hbar}[H_{I,m}, \rho_q]+\gamma_q{\cal D}[\sigma_-]\rho_q
\end{equation}
The signal on the transmission line is ultimately subjected to a phase-sensitive measurement such as homodyne detection. We will assume this is ideal and the phase reference is chose appropriately. The homodyne current is a stochastic process that satisfies the Ito stochastic differential equation\cite{WisMil}
\begin{equation}
J_j(t)dt =\langle Y_j\rangle _cdt+\frac{1}{\sqrt{\gamma_q}}dW(t)
\end{equation}
where $\langle Y_j\rangle_c$ is a conditional average of the Pauli y-operator for each qubit conditioned on the quantum state up to time time give the entire measurement record, and $dW(t)$ is the Wiener increment.  The conditional state obeys the stochastic Schr\"{o}dinger equation,
\begin{equation}
d\rho_{q,c} = -\frac{i}{\hbar}[H_{I,m}, \rho_{q,c}]dt+\gamma_q{\cal D}[\sigma_-]\rho_{q,c}dt+\sqrt{\gamma_q}dW(t){\cal H}[\sigma_y]\rho_{q,c}
\end{equation}
where the nonlinear super operator is defined by 
\begin{equation}
{\cal H}[A]\rho= A\rho+\rho A^\dagger -{\rm tr}[(A+A^\dagger)\rho]
\end{equation}

As we want the measurement to be as fast as possible, we will assume that the qubit is rapidly damped and adiabatically eliminate it from the total irreversible dynamics, see \cite{Gangat} for a similar example. 
We then find that the master equation for the cavity fields acquires an extra them of the form
\begin{equation}
\frac{d\rho}{dt} = {\cal L}_+\rho-\Lambda[\hat{n}_1,[\hat{n}_1,\rho]]
\end{equation}
where the number decoherence rate $\Lambda$ is given by 
\begin{equation}
\Lambda =\frac{4\kappa^2}{\gamma_q}
\end{equation}

 The measured signal, after adiabatic elimination (and rescaling) then obeys
 \begin{equation}
 \label{number-record}
M(t)dt =\langle a_1^\dagger a_1\rangle _cdt+\frac{1}{\sqrt{\Lambda}}dW(t)
\end{equation}
Where the conditional state of the two cavities now obeys
\begin{equation}
d\rho_c = {\cal L}_+\rho_c dt-\Lambda[\hat{n}_1,[\hat{n}_1,\rho_c]]dt+\sqrt{\Lambda}dW(t){\cal H}[\hat{n}_1]\rho_{c}
\end{equation}
Note that if each cavity starts in a thermal state the density operator stays diagonal in number. We can thus replace the quantum 
conditional ME above with an equivalent classical conditional master equation. This  makes simulations easier. 

We can now simulate the stochastic measurement record as in \cite{Gangat}. In the good measurement limit  we find that $M(T)$
is a multi-valued random telegraph process which takes values in the positive integers. The jumps up and down in the record correspond to 
transitions in the photon number in cavity one due to the Poisson processes $dN_{12}$ and $dN_{21}$ respectively. 

If we  average the stochastic measurement records in Eq. (\ref{number-record}) for an ensemble of trials 
we see that the average signal $s(t) ={\cal E}[M(t)]$ is simply the unconditional mean photon number in cavity one.   
\begin{equation}
s(t) = \langle \hat{n}_1\rangle
\end{equation}
In the ensemble average we can see the model operating as a kind of Mach clock. But we can do better than this. We can treat a particular 
stochastic record of jumps as an irreversible clock  in analogy with radio carbon dating.


\begin{references}
\bibitem{fn:allClocksIrr}The Weisenheimer amongst us might remark that in fact all clocks are irreversible since damping is ubiquitous.  Well, yes, however we hope the reader appreciates the distinction between a clock whose operation is based on reversible principles, and one that is based on irreversible.
\end{references}
\end{document}